\documentclass{vgtc}                          

\ifpdf
  \pdfoutput=1\relax                   
  \usepackage{graphicx}                
  \DeclareGraphicsExtensions{.pdf,.png,.jpg,.jpeg} 
\else
  \usepackage{graphicx}                
  \DeclareGraphicsExtensions{.eps}     
\fi%

\graphicspath{{figures/}{./}} 

\usepackage{multicol}
\usepackage{caption}
\usepackage{subcaption}
\DeclareCaptionSubType[Alph]{figure}

\usepackage{flushend}

\usepackage{microtype}                 
\PassOptionsToPackage{warn}{textcomp}  
\usepackage{textcomp}                  
\usepackage{mathptmx}                  
\usepackage{times}                     
\usepackage{cite}                      
\usepackage{tabu}                      
\usepackage{booktabs}                  
\usepackage{hyperref}
\usepackage{enumitem}

\usepackage[table]{xcolor}
\newcommand{\added}[1]{\textcolor{black}{#1}}

\onlineid{0}

\vgtccategory{Research}

\title{A Visual Analytics Framework for Adversarial Text Generation}

\author{Brandon Laughlin\thanks{e-mail: Brandon.Laughlin@uoit.net}\\ %
     \scriptsize Ontario Tech University %
\and Christopher Collins\thanks{e-mail: Christopher.Collins@uoit.ca}\\ %
     \scriptsize Ontario Tech University %
\and Karthik Sankaranarayanan\thanks{e-mail: Karthik.Sankaranarayanan@uoit.ca}\\ %
     \scriptsize Ontario Tech Univeristy %
\and Khalil El-Khatib\thanks{e-mail: Khalil.El-Khatib@uoit.ca}\\ %
     \scriptsize Ontario Tech University}

\abstract{\added{This paper presents a framework which enables a user to more easily make corrections to adversarial texts. While attack algorithms have been demonstrated to automatically build adversaries, changes made by the algorithms can often have poor semantics or syntax. Our framework is designed to facilitate human intervention by aiding users in making corrections.  The framework extends existing attack algorithms to work within an evolutionary attack process paired with a visual analytics loop. Using an interactive dashboard a user is able to review the generation process in real time and receive suggestions from the system for edits to be made.  The adversaries can be used to both diagnose robustness issues within a single classifier or to compare various classifier options. With the weaknesses identified, the framework can also be used as a first step in mitigating adversarial threats.  The framework can be used as part of further research into defense methods in which the adversarial examples are used to evaluate new countermeasures.  We demonstrate the framework with a word swapping attack for the task of sentiment classification.} %
} 


\CCScatlist{
  \CCScatTwelve{[Human-centered computing]}{}{}{Visual analytics};
  \CCScatTwelve{[Security and privacy]}{}{}{Spoofing attacks}
}

\teaser{
  \centering
  \includegraphics[width=\linewidth]{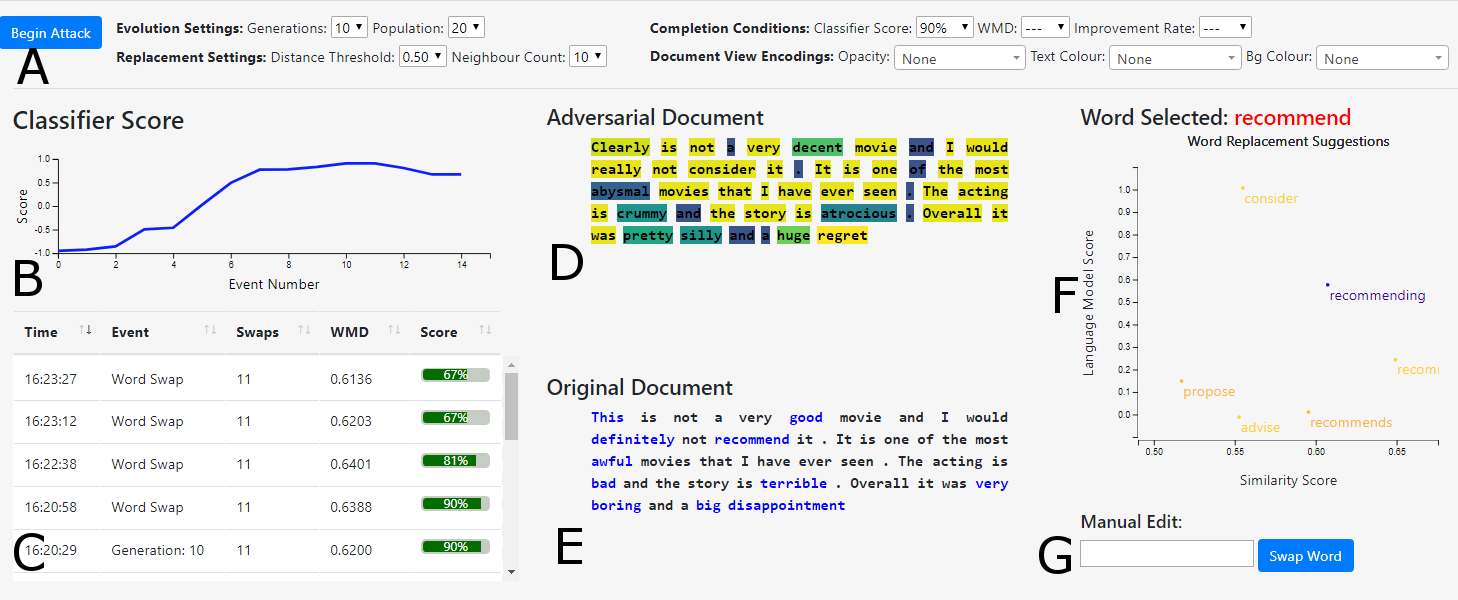}
  \caption{A linked visualization dashboard comprised of seven main components:  The settings area (A), a line chart for tracking the completion condition (B), an event log of previous interactions (C), document views for  the adversarial text (D) and original text (E), a scatterplot to suggest alternative word replacements (F), a text field for manual word substitution (G).  }
	\label{fig:teaser}
}

\begin{document}


\firstsection{Introduction}

\maketitle

\added{With the increasing challenges of securing big data, machine learning has become a popular tool for security applications. Adversarial machine learning is a threat to classifiers where malicious actors craft inputs with nearly identical features to the original data while being assigned a different output class. This work examines this threat for text classification where many automated attacks algorithms have been successfully demonstrated \cite{Zhang2019}.} In recent years with advances in techniques including deep neural networks and transfer learning, performance of many natural language processing (NLP) has been rapidly improving \cite{nlpprogress2019}.  The methods used in these state-of-the-art results however assume that the classifier input has not been manipulated \cite{liu2018survey}. With adversarial machine learning there is the potential for situations in which input data can be purposely generated by an adversary that wishes to manipulate the results of a classifier.  By using data from outside the trained distribution, malicious users can exploit the system by changing the assigned output class without significantly changing the content.  

Beginning with adversarial examples for computer vision \cite{szegedy2013intriguing}, the increasing popularity of deep learning has brought more attention to the susceptibility of deep learning models to these attacks. While adversarial environments have been studied for some time \cite{Biggio2018}, it is only recently with these increasingly complex classifiers that the issue has become more prevalent.  While the majority of works have centered around computer vision \cite{akhtar2018threat}, recently more works have been considering NLP \cite{Zhang2019}.  Generating adversaries for text brings additional challenges not found in the continuous data space of computer vision.  For example, the common evasion technique for computer vision is to slightly change each feature (pixel) by a small amount.  Each pixel can have its colours shifted very slightly without being noticed by a human \cite{akhtar2018threat}. This is more challenging for NLP as the data features are discrete.  One cannot simply alter a word slightly without being noticed.  Either a misspelled word or an entirely new word must be swapped with the original.  This means that instead of moving all words slightly one needs to choose specific words that will be modified.  While words can be converted to vectors in a continuous space, a slight shift in the vector space is unlikely to land on another word \cite{Zhang2019}. 

Due to the large vocabulary space of languages, the same idea can be expressed in many ways, providing the flexibility to construct alternate phrasings. With this ability the goal \added {of the adversary} can be to change a text so that the target class is changed but a human would still obtain the same meaning as the original document.  When developing NLP applications, replacing a word to a word with a very different meaning should alter the classification. However with adversarial examples, score changes can occur even when swapping semantically equivalent words~\cite{Alzantot2018}. While this can be done entirely through an automated attack algorithm, we advocate semi-automation where a human remains in the loop. This is important for the generation of adversarial texts as semantics are largely context dependent and therefore in need of manual user review. 

Trying to troubleshoot the reasons for a lack of model robustness is complicated because of poor model interpretability. Deep learning classifiers essentially act as a black box with seemingly random reasoning as to the decision of the model \cite{adadi2018peeking}. Due to this it can be hard to determine how a change to the input will influence the output. Recently there has been a lot of attention on creating methods of better explaining complex machine learning systems \cite{mittelstadt2018explaining}.  It has become a popular research topic with initiatives such as Explainable Artificial Intelligence (XAI) \cite{DavidGunning2017} that has the objective of exploring  methods for increasing transparency in AI systems.

We propose a framework that combines an attack algorithm and a visual analytics dashboard for a human centered approach to generating adversarial texts. Starting with an automated evolutionary attack process, the system builds an adversarial example over a set of generations.  After the algorithm has completed, the user can use an interactive dashboard to make adjustments to the resulting text to correct the semantics.  The end objective is a more automated and efficient way to craft adversarial examples while allowing the user to adjust poor changes by the attack algorithm. Since the system uses an approach that is both black box and model agnostic, it is flexible and can be transferred to other classification tasks. The provided example in this paper is classifying document sentiment.

From the perspective of an attacker, adversarial examples can be used for tasks such as spamming, phishing attacks, online harassment and the spread of misleading information. This framework could be used as a way to combat such malicious activities through several uses. \added{First, the framework identifies adversarial weaknesses for classifiers, by reviewing actual attack instances a user can better diagnose vulnerabilities in their classifiers.  Once threats have been discovered users can correct any semantic inconsistencies in the examples and test whether the threats persist.  Another benefit of crafting adversaries is using these semantically equivalent adversaries for adversarial training \cite{goodfellow2014explaining} to strengthen the classifier.  When a user has multiple model options they can do attack testing to compare the robustness of different models and choose the best one. Building a set of semantically equivalent adversaries can help to design more realistic and diverse threat models with varying levels of difficulty.  Lastly our work can be used by security researchers in research on adversarial defense.  Used as a means for building high quality training sets, the resulting adversaries can be used in research for new defense techniques.}

\added{Our contribution is as follows:}
\begin{itemize}[leftmargin=*,itemsep=0.3ex]
    \item \added{A visual analytics framework which extends existing attack approaches to visualize specific attack instances in real time.}
    
    \item \added{Visualizations to support users in crafting semantically appropriate adversaries. A document view highlighting semantic inconsistencies is paired with a scatterplot that suggests word replacements.}
    
    \item \added{A risk management tool for testing different threat models. Using different text manipulation methods a user can compare results across different classifiers.}

\end{itemize}

\section{Background and Related Works}

Our framework involves research into the robustness and interpretability of machine learning models and the ways humans can be involved to improve results.  In this section we start with a review of related works on adversarial machine learning, followed by research on how visual analytics has been used to help address these issues.

\subsection{Adversarial Machine Learning}

There are two main forms of adversarial attacks: white box and black box.  White box attacks involve having access to information about a classifier including the training data, the features, the scoring function, the model weights and the hyper parameters \cite{Biggio2018}.  Black box attacks do not require any of this information and is the approach our framework supports.  The only feedback required is the classifier score. Another distinction in attack types is the source from which an adversary can inject samples.  The two options available are poisoning and evading \cite{Biggio2018}.  Poisoning involves the ability to place adversarial examples in the training data that then becomes part of the data distribution the model is trained on.  Evasion attacks only have access to a model after training has been complete.  Our work deals with evasion as our task is to cause mistakes in classifiers that would already be trained. In instances where the classifier is updated over time on future iterations of data, some evasion samples may be part of future training, in part achieving a poisoning attack.

The underlying issue that enables adversarial examples to pose a threat to machine learning classifiers is a lack of robustness \cite{liu2018survey}.  We define robustness as the degree to which a change in the input of the classifier results in a change to the output score.  A robust model is more stable and therefore more predictable in the scores it generates.  A less robust model can have a drastically different score for seemingly very similar inputs. It has been found that NLP classifiers often provide unrealistic confidence estimates that result in seemingly random inputs having high probability scores \cite{Feng2018}.  This was tested by using input reduction that involves iteratively removing a word from a document until the classification score changes greatly \cite{Feng2018}. The authors found that all of the relevant words can be removed while leaving seemingly random words to a human observer.  The solutions end up as several nonsensical words unrelated to the original document. This demonstrates the risks involved with model over-fitting \added{(the loss of a model's ability to generalize)}.

\begin{figure*}[tb]
 \centering 
 \includegraphics[width=\textwidth]{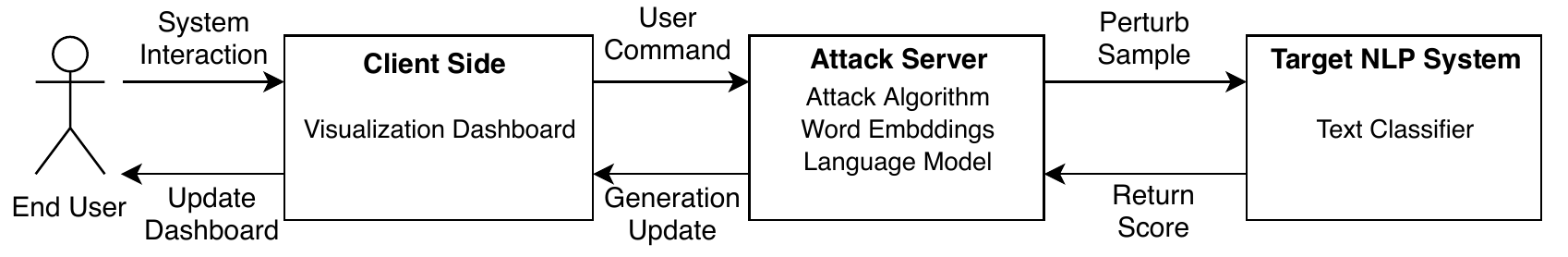}
 \caption{Diagram of the framework architecture.  The user interacts with the dashboard which sends commands to the server.  The attack algorithm will send samples to the target classifier for testing.  The resulting scores are processed by the attack system and an update is sent to the dashboard.}
 \vspace{-4mm}
 \label{fig:framework_overview}
\end{figure*}

The most common adversarial attack method in NLP is the white box scenario with this attack type more easily able to manipulate examples \cite{Zhang2019}.  Black box attacks are more challenging but have still found success. \added{As detailed further in Section \ref{Attack Selection}, our framework is designed to work with any black box attack.  Examples below are of different perturbation strategies that could be adapted to work with our evolutionary attack system.  While these works automatically generate all of the adversaries, our framework is designed to extend these strategies to work with human intervention in a visual analytics dashboard.}    Attack types can be divided into two main groups: character based and word based.  Character level attacks perform perturbations on individual characters within a word.  Basic examples include attacks that use insertion, deletion, swapping, and substitution \cite{Li2019}.  More advanced versions can use rare characters such as from other languages that are similar visually to the original character \cite{Eger2019}.  Word level attacks involve modifications at the level of entire words, most often replacing it with a semantically similar one.  One example is replacement using neural machine translation \cite{Ribeiro2018}.  The examples demonstrated in our work are based on a technique that swaps words using a genetic algorithm \cite{Alzantot2018}.

\subsection{Visual Analytics}
There have been many works that involve the use of interactive visualizations for the explanation of machine learning models \cite{liu2017towards}. However, visual analytics work on adversarial attacks are limited. Related work pertaining to visualization of adversarial attacks is Adversarial Playground \cite{norton2017adversarial} which specifically works with white box attacks for computer vision. Our work evaluates the use of adversarial examples in the discrete domain of NLP using black box and model agnostic approaches. \added{  In the domain of NLP a tool has been built to help users craft questions for a quiz game that trick information retrieval models \cite{wallace2018trick}.  The tool shows evidence for or against the right answer by highlighting influential words. Our work differs as our tool is designed for any text classification task while their tool works specifically for question answering algorithms.}  To our knowledge there does not yet exist any visual analytics research exploring the generation of adversaries for NLP classification.  

Visual analytics systems have been designed to help users examine machine learning classifiers to help improve classifier performance though methods such as explanatory debugging \cite{kulesza2015principles}.  Providing explanations for classifier errors has been found to improve trust in the predictions made by the model \cite{dzindolet2003role}.
Our framework employs similar techniques by explaining behaviour with scores built from embedding distances and language models.  In our task we have the specific debugging objective of making semantic corrections to the output by improving word swaps made by the attack algorithm. 

Similar to our framework, other works have used visualizations to explain the behavior of classifiers. To explain model decisions, feature heat maps have been built that colour encode the words based on their importance to the classifier score \cite{Feng2018}.  Other works have visualized partial dependence plots \cite{krause2016interacting} and provided instance-level explanations \cite{tamagnini2017interpreting}. Interpretability for a model can be defined as a global or local process \cite{Molnar2018}.  A global explanation would provide details of how the system works as a whole to generate all of the results.  Local scope explanations provide context to specific subsets or individual instances of the dataset, such as LIME \cite{ribeiro2016should}. Our framework provides local explanations for individual word selections.  This means that the impact of word replacements are calculated at that specific location with the surrounding context taken into consideration with language models.

\added{Most NLP classifiers represent the input using word embeddings which are numeric representations of words in a high dimensional vector space.  Words with similar semantic meaning are positioned more closely in this vector space compared to words representing less related concepts.  In order to visualize the relationship of words we represent the embedding relations using the euclidean and cosine distance between the word vectors.}  To help users more easily find word replacements, we provide scatterplots to explore the word embedding space. Related works have that have explored embedding spaces include a tool for concept building \cite{park2017conceptvector} as well as comparing word relationships and embedding spaces \cite{heimerl2018interactive}.

\section{Framework Description}

The framework is designed as a combination of a client facing web dashboard and an attack server on the backend.  The server contains the attack algorithms, word embeddings and language model.  The user interacts with the server through the web dashboard which translates the interactions into commands for the server.  The server then attacks the target classifier using an attack configuration chosen by the user.  An overview of the framework can be seen in Figure~\ref{fig:framework_overview}.  This architecture design was chosen to easily facilitate communication between the client and the attack process.  The analytics can be done on a powerful server that can deliver the results to a web browser on any device.  The web browser enables an interactive visualization dashboard to present the results of the service to the user. 

In order to support uses in a wide variety of environments and use cases, flexibility was a central design goal. To support a flexible approach, the framework is model-agnostic and supports any black box attack. All parts of the architecture are delivered as a black box where the user does not need to know the underlying details of any component used as they are abstracted into the service. There are many benefits to model-agnostic systems including model flexibility and representation flexibility \cite{Ribeiro2016}.  Model flexibility means that the system must be able to work with any machine learning algorithm and this ensures that our attack algorithm will work against any type of classifier.  As an example whether the target classifier being attacked is a rule-based system, a neural network or any other classifier the attack will work in the same way.  The only requirement from the classifier being attacked is that the assigned class needs to come with a numerical score. Representation flexibility means that the system supports many explanation methods. Having different explanations can help the user adjust to different objectives and domains.  Our framework supports such flexibility by allowing the user to easily switch in different word embeddings and language models.

\subsection{Attack Selection} \label{Attack Selection}
When generating adversarial texts there are many factors to consider that impact the quality of the resulting examples.   These constraints can be described as the attacker's action space and describe the set of constraints the adversarial examples must meet \cite{Gilmer2018}.  They are some of the considerations needed when defining how an attack should operate.   The following is a list of what we consider to be some of the most important factors to consider when defining an attack:
\begin{itemize}[leftmargin=*,itemsep=0.3ex]
\item \textbf{Content-preserving:} The text must preserve the content of the message.  For example, if the text is about a particular named entity, it must still be about that entity. 
\item \textbf{Semantic-preserving:} The attacker may make any perturbation to the example they want, as long as the meaning is preserved.  
\item \textbf{Syntax-preserving:} The grammatical elements of the text should be the same, the structure of the writing should remain unchanged.
\item \textbf{Suspicion:} To what extent the text appears to be purposely manipulated. An example would be replacing characters with alternative symbols.
\item \textbf{Legibility:} The text is in a form that can be read by humans.  For example, visually perturbing text such as through \textit{captcha} techniques would degrade legibility.
\item \textbf{Readability:}  The text can still be easily understood by a human.  For example, replacing text with words beyond the average person's lexicon would degrade readability.
\end{itemize}

The extent to which an attack matches the above criteria is often a subjective question and therefore is likely to be placed somewhere on a spectrum for each of these aspects.  For example, spelling errors or poor grammar might increase suspicion but how much is uncertain as these could be considered legitimate mistakes.  This could also possibly impair syntax, semantics or readability. Depending on the importance of the various constraints, different attack strategies need to be implemented. The framework is designed to allow the user to use many attacks types so that these constraints can be considered.  

With all of these different constraints there might be multiple attack options to decide between when choosing an attack strategy. An attack agnostic framework makes comparing and switching between options more easy.  It may be difficult to compare the effectiveness for two attacks such as a character-based versus word-based attack.  With our attack-agnostic system, both options can be fed into the system and provided the same type of assessment. With the same representation used, a direct comparison becomes more easy.  Additionally the flexibilities afforded by an attack-agnostic system offers the ability to switch between them more easily.  During the same attack a user could switch attack strategies. The demonstration in this paper assumes that the resulting text must not be suspicious and so we use a word swapping approach. Semantics, syntax and content might be still be impaired; this is why the user is involved in making appropriate adjustments with the dashboard. 

The attacks are implemented as a genetic algorithm which emulates the process of natural selection.  This is done by iteratively building stronger generations of adversarial texts over time. The solutions evolve through crossover from parent reproduction and mutations of the text. The purpose of the mutations is to add diversity to the documents to more effectively explore the search space.  Reproduction is used to increase the propagation of favourable outcomes and reduce the frequency of poor performing documents. The likelihood of each parent reproducing is determined by a fitness score.  The fitness score is based on the output score generated from the target classifier we are attacking.  The fitness score improves as the output score gets closer to the target class.  The specific conditions for mutation and reproduction vary by attack strategy.  The word swap approach demonstrated in this work is detailed in the use case section.
  
When an attack has been selected the user then chooses the evolutionary parameters including the number of generations, the population size and the word swap settings for how many nearest neighbours to return and a cutoff threshold based on the distance in the embedding space.  With the attack chosen and parameters set the last step is to define the completion conditions. \added{The completion conditions define at what point the system will stop attacking each document.  Once any of the defined conditions have been met the evolutionary process will end and move on to the next document.  If no conditions are met then by default the attack will continue until the maximum number of generations has been reached.} The following is a list of completion conditions that can be set:

\begin{itemize}[leftmargin=*,itemsep=0.3ex]
\item \textbf{Classifier Score:} In situations where the text needs to reach a specific classification score such as passing through a filtering system.  When the score passes this threshold such as negative to positive (above zero) the process will stop. 
\item \textbf{Word Mover's Distance \cite{Kusner2015}:} Can be used as a way to roughly approximate the overall extent of change between the current adversary and the original document.
\item \textbf{Duration:} Once the attack has continued past a specified amount of time, end the attack after the next generation completes.
\item \textbf{Performance Acceleration:} Once the rate of improvement in classifier score between generations drops past a specified threshold the attack ends.
\end{itemize}

\begin{figure}[tb]
 \centering 
 \includegraphics[width=0.75\columnwidth]{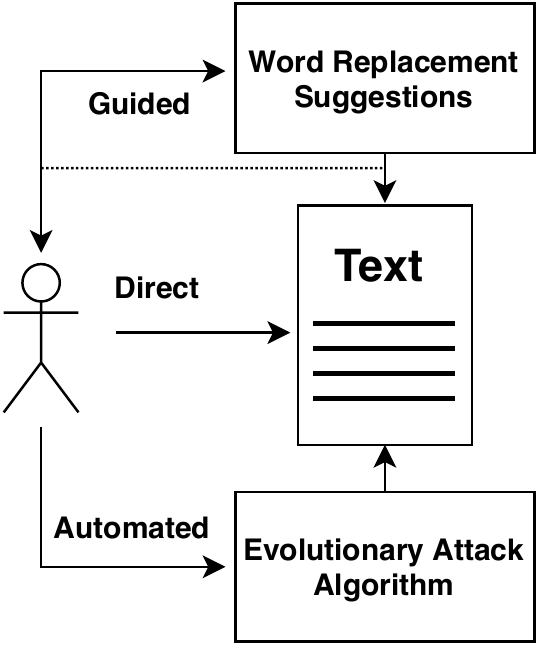}
 \caption{The user can interact with the attack generation through guided word replacement, through direct editing, or by allowing the evolutionary attack to run automatically.}
  \vspace{-5mm}
 \label{fig:Interaction_Methods}
\end{figure}

\subsection{Interaction Methods}
As the framework is built for black box evasion attacks, the attack process involves repeatedly sending slightly perturbed inputs until the target class is reached.  For this reason an automated approach is virtually essential as it would be extremely time consuming for a human to repeatedly craft new inputs. Additionally, when humans test subjects are given the task of creating adversarial texts they have difficulty coming up with examples.  When automated approaches were tested they were found to be much better at this task \cite{Ribeiro2018}.

However, while the algorithms are good at generating candidate solutions, they are unable to always make the best decisions.  While humans cannot build examples easily they have a skillset complementary to the machine which is to easily select the best text among several options \cite{Ribeiro2018}.  Therefore, while some form of automation is needed it is important to have the user involved in the process. The combination of both human and machine can outperform just the attack algorithm alone.  Human intervention is also needed because of the complexities of human language.  At least some user feedback is needed to guide the algorithms as context plays a large role in text analysis.  Since word similarity is very dependent on the context of surrounding words we still need to rely on a human for final review.  Adjustments are needed in situations where the classifier has chosen words that change the semantics of the text.

The assumption when working with word embeddings is that the nearest neighbours of the words will be the ones that are the most similar semantically.  However this does not always ensure that true synonyms will be the nearest neighbours.  For example, antonyms and hypernyms can be found close in the embedding space as they are used in similar contexts as the original word.  In this work we prevent stop words from being swapped  and filter the results through WordNet~\cite{miller1995wordnet} to check for antonyms close in the embedding space.  These additions, however, do not  guarantee nearest neighbours are truly similar.  Even when words are proper synonyms, other challenges such as words with multiple meanings complicate the simple word swapping approach. It is for this reason that the attack has been integrated with a human-centered visual analytics dashboard to allow the user to make changes as needed.  The automation of the attack algorithm needs to be combined with the subjective insights of a human user. 

For these reasons the framework supports three methods of interaction: using the evolutionary attack algorithm (automated), with nearest neighbour exploration (guided) and manually through the text form view (direct).  These options can be seen in Figure \ref{fig:Interaction_Methods}. The suggested interaction order is to first run an automated evolutionary attack followed by guided scatterplot suggestions. If there are still issues with the text then the user can directly edit the text.  The user is free to use any combination of these methods and in any order. The manual and guided interactions offer direct manipulation of the system without having to launch an entire attack.  This can be useful to test a quick hypothesis or troubleshoot the system. The user can also begin another automated stage using the evolutionary attack using the current edits as a new starting position. To prevent the algorithm from simply switching back the words the user has changed, any words edited by the user are automatically locked.  This lock prevents the algorithm from making any further changes to these words.  

\section{Dashboard Description}
The combination of an automated algorithm working together with a human analyst provides a good opportunity to use visual analytics to more easily integrate together the two parts. As seen in Figure~\ref{fig:teaser} the dashboard is organized into seven parts.  All of these components are connected together in a single linked visualization dashboard.  Across the top is the attack configuration settings (A). Below this the line chart (B) tracks the score of the classifier or any other completion condition. Below this an interactive table logs progress and enables the loading of previous snapshots (C). The center displays the adversarial (D) and original (E) documents.  On the right is the scatterplot view (F) for selecting word replacements. Manual word replacements can be done with the input field (G). When an attack has been started, the algorithm iterates through all of the generations and provides an update after each generation is complete.  The best example from each generation is known as the \textit{elite} and is used to represent the progress of the attack.  The server updates the dashboard with this elite.  The document view, the linechart and the event log are updated for each generation in real time as the attack progresses.


\begin{figure}[tb]
\begin{minipage}[b]{\columnwidth}
\includegraphics[width=\columnwidth]{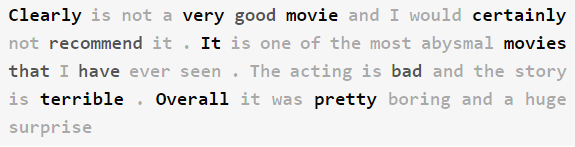}
\subcaption{Classifier score importance as text opacity.}
\end{minipage}
\begin{minipage}[b]{\columnwidth}
\includegraphics[width=\columnwidth]{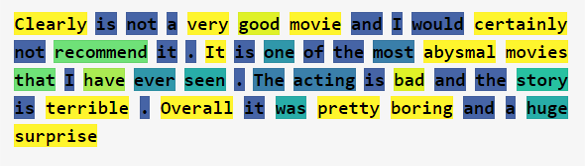}
\subcaption{Word selection probability as background color.}
\end{minipage}
\begin{minipage}[b]{\columnwidth}
\includegraphics[width=\columnwidth]{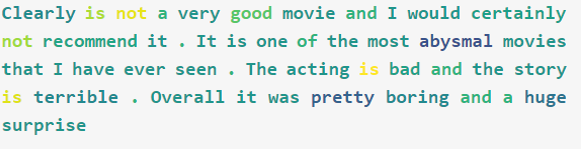}
\subcaption{Language model score as text color.}
\end{minipage}
\begin{minipage}[b]{\columnwidth}
\includegraphics[width=\columnwidth]{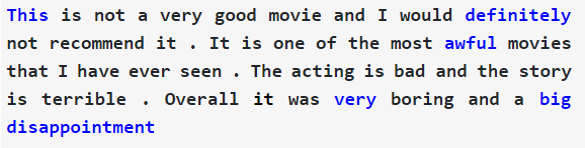}
\subcaption{Original document with swapped words colored blue.}
\end{minipage}
\caption{Document view encoding options. The scoring options (A-C) can be paired with the original document (D).}
 \vspace{-5mm}
 \label{fig:document_view}
\end{figure}

\subsection{Document View}
The document view shows the current state of the adversarial example (Figure \ref{fig:teaser}D) as well as the original text (Figure \ref{fig:teaser}E).  While the attack algorithm is running, this view is updated to display the best example (elite) from each generation.  Once the attack algorithm has been completed, the final adversary is presented to the user. The words within the adversarial document can be visually encoded according to several objectives: classifier score influence, word selection probability or semantic quality. 

The words that have been changed between the original and adversary are coloured blue in the original document for quick identification. \added {An older design also coloured the swapped words in the adversary text.  This however would add an additional colour element to the adversary document encodings described below which would likely be overwhelming for a user.  For this reason we decided to have the same position for each word within the adversarial and original texts linked together.} When the user hovers over each word, both words in that same position are highlighted.  This allows the user to easily orient themselves with the same word position in both texts at once

The score encoding shows the impact the words have on the classifier score. The score is calculated for every word by replacing each word with the word's nearest neighbour in the embedding space.  The document is scored before and after the word has been swapped. The final score is the difference in score from the first word and the new swapped word.  For example, if the current word is `bad', the nearest neighbour `terrible' is put in that position instead.  Two instances of the document are then scored, one for 'bad' and once as `terrible'.  The difference in scores is kept and are represented as the opacity of the original word.  This comparison provides a rough approximation as to the importance of each word and lets the user easily spot good candidates for score improvements.  If the swap improved the score it would be given a higher opacity and a reduced score would have a lower opacity.  As seen Figure \ref{fig:document_view}A, words such as `Clearly' and `certainly' would be good candidates whereas words such as `acting' and `story' would not.

The word selection encoding represents the probability of each word being chosen by the attack algorithm. This is based on a count of how many nearest neighbours each word has.  This is calculated as the number of words nearby in the embedding space within the threshold specified by the user. The number of nearby words is converted to a probability based on the relative word counts of the other words in the document.  Words that share a similar meaning to many other words are likely to have a much higher count than more unusual words.  This view can enable a user to quickly see which words are more likely to have suitable replacement suggestions available by the system without having to load the scatterplot for each word.  The background colour of the text is used to represent the probability using the Viridis blue-yellow color scale.  Words with a higher probability are given a more yellow (brighter) background colour. As seen in Figure \ref{fig:document_view}B, `very' has many options, `recommend' has some options and `acting' has very few.

The last document view option is the semantic perspective that visually encodes the words according to their probability score from a language model.  This view can be used when the user wants to improve the semantics of the text. Each word is processed by the language model with its surrounding context to determine a probability score that reflects how appropriate each word is in that spot.  This view can help a user identify words that are not appropriate for the sentence and that need to be changed. The brightness of the text colour is used to represent the semantic score. Here, lower semantic score is more blue, which through luminance contrast with the background drives attention to words that are better candidates for editing. As seen in Figure \ref{fig:document_view}C, the majority of the words have an average score and the word `abysmal' is one of the least appropriate.

The user can choose to represent one or any combination of these encodings at any time. Once the word encodings have been calculated the user can begin to select individual words to swap. The user can activate any word from the text by clicking on it.  This word now appears in the top right corner of the dashboard and the word replacement view is activated with this word which is described further in the next subsection.  To help the user easily identify the selected word within the text, the selected word is given a dark background within the document text. Whenever the user swaps a word the document view is updated. Each word in the text is again scored by the classifier and then the encodings are updated. 

\begin{figure*}[tb]
 \includegraphics[width=\textwidth]{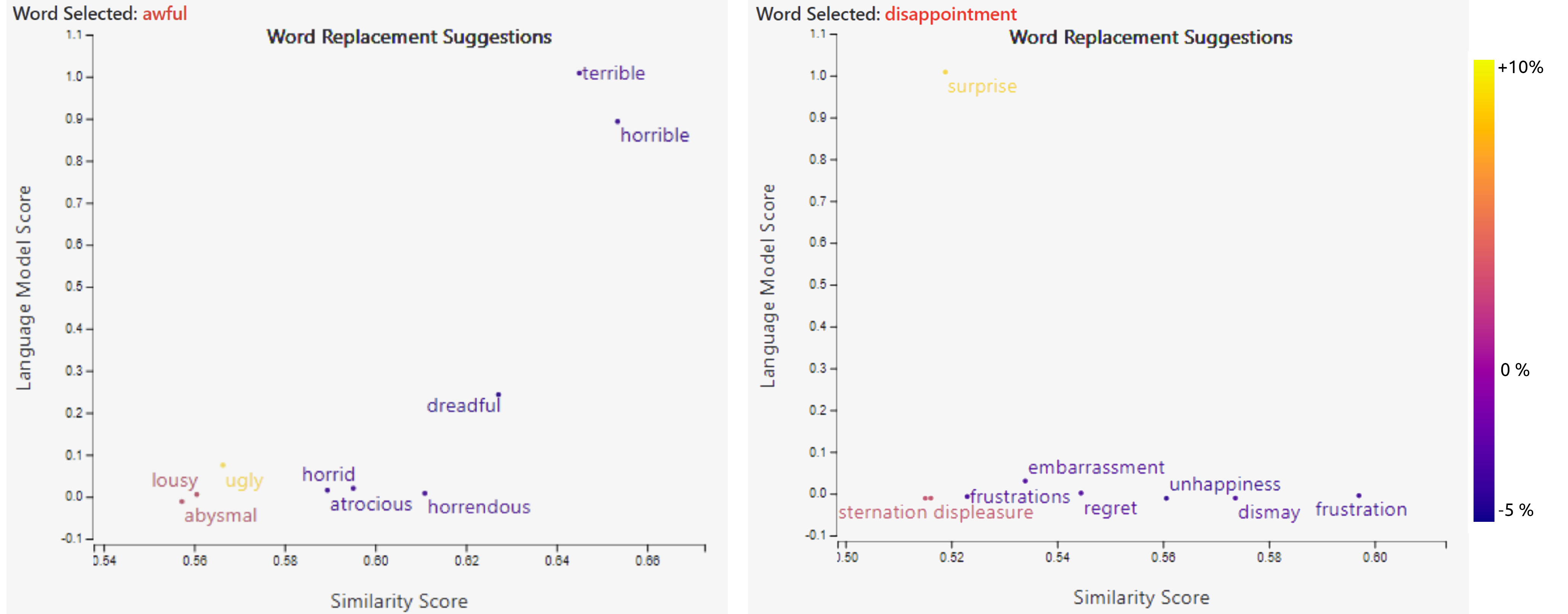}
 \caption{Two example scatterplots of word replacement options. Y-axis plots probability from a language model, x-axis plots similarity score based on word embedding distance, and hue encodes change in classifier score. Left: replacement suggestions for the word `awful'. Right: replacement suggestions for the word `disappointment'.}
  \vspace{-5mm}
 \label{fig:scatterplots}
\end{figure*}

\subsection{Word Replacements}\label{Word Replacements}
The word replacement section is the right side of the dashboard and is where the user can choose word replacements with either the scatterplot suggestions (Figure \ref{fig:teaser}F) or manually with the text field (Figure \ref{fig:teaser}G). The scatterplot view is used as a guided interaction to help users more easily identify suitable word replacements. The purpose of the scatterplot is to see what would happen if any of the nearest neighbour candidates was chosen to replace the current word in the adversarial text instead. This enables the user to quickly find any appropriate replacements for the current word selected.

When a user selects a word the attack server retrieves all of the nearest neighbours of that word within the defined distance threshold by using the word embedding space. For each of the nearest neighbours three scores are computed: a classifier score from the model we are attacking, a probability from our language model and a similarity score.   The classifier score is calculated as the difference in score between this word and the original word (the word that was clicked on). These score encodings function the same way as classifier scores in the document view. That is, each word is compared to the current word in the text by replacing it in the document and running it through the classifier. The embedding space similarity score for each candidate word is computed based on the embedding distance to the current word in the text. The similarity scores range from 0 to 1 and our implementation is based on the Euclidean distance in the Google News corpus word2vec embedding \cite{mikolov2013efficient}. Larger numbers indicate more similarity between this word and the current word. The language model probability scores are compared between all the replacement candidates. For the language model we use the Google 1 billion words model \cite{chelba2013one}. Words that fit most appropriately in the surrounding context will have larger scores than those that do not. The scores are normalized between the range of 0 and 1.

Once all of the words have been retrieved and their scores computed, they are placed on the scatterplot. The x-axis plots the similarity score and the y-axis plots the language model probabilities. The axis for both of the scatterplot features starts at zero and increases towards one. This means words near the origin are the least desirable for both features. Farther out upwards and towards the right improves the semantic score and the similarity of the words respectively. The colour brightness of the words is encoded with the classifier scores using the d3 plasma blue to yellow color scale. With all three features considered, a user would ideally find bright words in the top right corner indicating similar words that fit the context and that also boost classifier performance. When a suitable word is found the user can click on that word to use it as a replacement.  The new selected word now takes the place of the old one and the document view updates.  

\added{An alternative design considered was giving the user the flexibility to choose which features to plot on the scatterplot axes.  For example, the change to the model score could be combined with one of the other semantic features.  We choose to keep the model scores as colour only to keep the design and use of the interface simple. Since the primary objective of the user is to improve semantics we believe that this extra option was not worth the increased complexity.  Additionally, keeping both positional encodings focused on semantics provides a clear distinction between the choice impact on semantics versus adversarial performance. }

Examples of scatterplots can be seen in Figure \ref{fig:scatterplots}. For `awful' (\ref{fig:scatterplots}-left) both the words `terrible' and `horrible' would be decent replacement options. They do however, reduce the classifier score which may render the options unusable if the classifier score is near the decision boundary. The options for  `disappointment' (\ref{fig:scatterplots}-right) are more disappointing as there are no clear winning candidates within the top right quadrant.  

The other human intervention method is to manually edit the text directly by using the text form that allows the user to edit the underlying text directly. The user may want to make manual edits if the word they want to use as a replacement is not suggested in the scatterplot.  Even if a word is in the scatterplot the exact version may not be appropriate and they may want to take a word and make some small adjustments such as editing the prefix or suffix of a word to more appropriately match the surrounding text. For instance the use may wish to make adjustments for issues such as proper word tense or switching between singular and plural versions of a word.  In these situations the user types in the desired word replacement into the text box and clicks the swap word button.  This achieves the same end as clicking a word in the scatterplot.

\subsection{Event Log}
The event log (Figure \ref{fig:teaser}C) is an interactive data table that records every action made by both the user and the attack algorithm.  For the algorithm, an update is sent after every evolutionary generation has completed.  For the user, any word replacements either by manual text edits or word swaps with the scatterplot view are added to the table.  For each action the following are recorded and stored in the table: a timestamp, an event description, the total swap count, the word mover's distance (WMD) \cite{Kusner2015} relative to the original document, and the score from the classifier. Each table column can be sorted by clicking on the column header.  The event log enables the user to review the impact on the document by sorting over time, interaction type or changes on the text (swaps, WMD, score).

When using non-linear classifiers, the user may wish to step through several interactions in a sequence to see if subsequent choices impact past decisions.  If a user wishes to revert any changes done they can do so through the data table log.  By clicking on any entry in the table they can return to this snapshot.  This allows users to easily revert back to previous decisions, allowing for non-permanent interactions.  This can more easily facilitate what if analysis by the user where they may wish to explore different options.

\section{Use Cases}
In this section we demonstrate an implementation of the attack algorithm and the process involved in adjusting an adversarial text using the dashboard. The end objective of the attack is to take a document which in this case is a negative movie review and convert this to a positive review without changing the semantics of the text. Since the review was originally negative, a human reading the review should still believe the review is negative even after it becomes classified positive by the machine learning model. The attack algorithm implemented in this work is based on an existing word swapping attack~\cite{Alzantot2018}.  For our evolutionary attack algorithm the mutations occur as words swaps for semantically similar words based on a word2vec embedding from the Google News corpus \cite{mikolov2013efficient}.  Nearest neighbour lists are built for each word in the document with a cut off over a specified distance in the embedding space.  The more neighbours a word has under the specified threshold the more likely it will be chosen as the mutation. Reproduction is implemented as crossovers involving two parents with a child randomly receiving half of the words from each parent.   

\subsection{Adversarial Dataset Building}
In this example the user wishes to construct adversarial examples in order to experiment with adversarial training~\cite{goodfellow2014explaining}.  With adversarial training the classifier is trained on adversarial examples in order to increase robustness against them. By using the attack algorithm alone the user might be training the model on adversaries that were actually not semantically similar to the original.  This would mean training would be done on improper adversarial examples so the results would not be as effective.  By making corrections to the texts with poor semantics, the training set quality for the adversarial training can be improved.

\begin{figure*}[tb]
 \includegraphics[width=\textwidth]{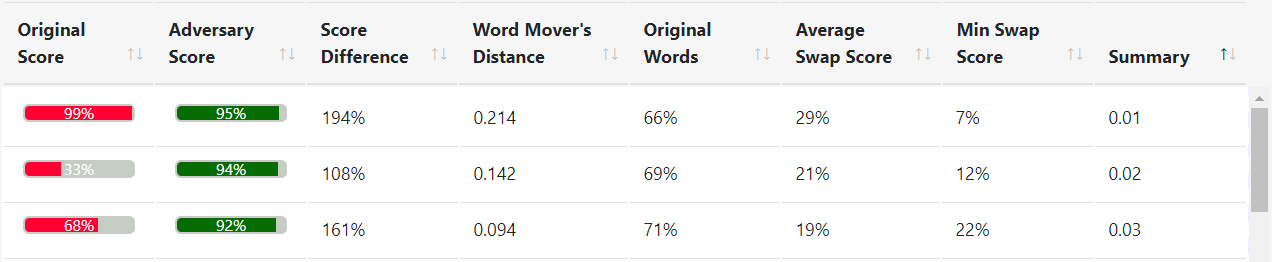}
 \caption{\added{A data table lists each of the adversarial texts generated.  Users can choose which text to edit based on metrics such as word mover's distance, the number of words swapped and the language model scores of the word swaps. }}
  \vspace{-5mm}
 \label{fig:adversary_selection}
\end{figure*}

To start the user selects the classifier score threshold as the completion constraint.  For our scenario the objective is to achieve a score of at least 0 (neutral). The user then generates many adversarial examples using different documents as the starting point, thus building a diverse set of adversarial examples. With an adversarial dataset built, the user needs to select which data records to investigate. 

\added{Figure \ref{fig:adversary_selection} shows an example of a data table which lists all the generated adversaries.  Options included in the table include the final word mover's distance (WMD) \cite{Kusner2015} of each document, the percentage of original words remaining and the average and minimum semantic scores of word swaps using the language model. A summary score can be defined to combine metrics in order to help users prioritize edits. In this instance it is the product of the original words remaining and the average and minimum swap scores.  A higher percentage is desirable for each category so a smaller product indicates a text with more potential corrections to be made.  The user chooses the example with the smallest summary score which opens the document in the interactive dashboard (Figure \ref{fig:teaser}).}

With a specific record now chosen the user will begin to examine the text and improve the semantics. The adversarial text chosen had successfully switched classes from negative to positive. The user now wants to confirm that the example is truly similar semantically to the original.  To quickly check for poor word substitutes that have been made, the user selects the language model encoding in the document view.  As seen in Figure \ref{fig:document_view}C the user sees that the word `abysmal'  which replaced `awful' has been identified as a word with a poor language model score.   The user also sees another replacement they wish to fix: `disappointment' has been replaced by `surprise'.  The user feels that there are more appropriate substitutes for these words. As discussed in Section \ref{Word Replacements}, the user selects both of these words within the document view and the results can be seen in Figure \ref{fig:scatterplots}.  The user chooses replacement words and repeats the process for each word they wish to correct. When the user does not wish to use any of the suggested replacements they insert their own word manually via the text field.  In instances where the user is unsatisfied with the change in scores, they can revert back to the previous snapshot using the event log. 

\begin{figure}[tb]
 \includegraphics[width=\columnwidth]{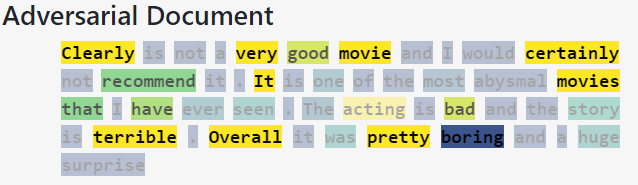}
 \caption{Adversarial document view with multiple encodings. Background brightness is word selection probability and text opacity is classifier score importance.}
  \vspace{-7mm}
 \label{fig:optimizing_adversary}
\end{figure}

The user continues to search for other words in the adversary to replace until all the poor semantic words have been fixed. The user has noticed that the classifier score has dropped beyond the threshold needed. They could launch another evolutionary attack or make changes themselves.  Since the score only needs a slight upgrade they decide to fix it themselves.  They search for the best opportunity for score changes by enabling the performance encoding to find words that have replacements that can improve the score.  They also add the word selection probability encoding to find words that are likely to have replacements. This can be seen in Figure~\ref{fig:optimizing_adversary}. They find the word `movie' has a good opportunity to increase the classifier score (high opacity) and has many suitable replacements (bright colour).  They then repeat the process of looking for replacements in the scatterplots. When the adversary has been fixed they can continue to search through other adversaries, returning to the data table in Figure \ref{fig:adversary_selection} and prioritize based on the summary scores.

\subsection {Attack Algorithm Adjustments}
The dashboard can also be used as a way to review the attack process of the evolutionary algorithm.  If used in this way the dashboard can be used as a troubleshooting tool to help debug errors or better optimize the attack results. To do this the user can change the encoding option in the document view to the word selection probability.  This will visualize the influence of each word during the attack to provide the user a better understanding of how the attack chooses which words to perturb. The development of the attack example can be reviewed after each evolution by looking at each generation of the attack using the event log.  By stepping through each stage the user can see which words are being replaced at any time in the attack.  The user can jump to a snapshot in the event log and bring up the document view.  

In this example the user wants to troubleshoot the attack algorithm.  Specifically they want to know why the word `It' has a high selection chance. As seen in Figure \ref{fig:optimizing_adversary} line 2, the user observes that the selection probability for the word `It' is very high which they find strange as they thought it was added to the list of stop words to ignore. The stop list is used for words in which there are no conceivable replacements as the word is uncommon or has no synonyms of any form. The user notices that other instances of the word `it' in this document were scored much lower, but then realizes that this one was at the start of the sentence so it was capitalized.  This capitalized version of the word was not part of the stop list. To fix this issue the user now adds this specific version of the word to the stop list.  Alternatively they could make the stop list case insensitive.

To improve the attack performance the user can look for words that have large discrepancies between the classifier score influence and the word selection probability.  That is, the user can look for words that have selection probabilities that do not properly reflect their importance to the classification. As an example in Figure \ref{fig:optimizing_adversary} bright, bold words are important to the score and likely to be changed.  Faded dark words are unimportant and unlikely to be chosen.  These are optimal conditions for the attack.  However the word `acting' (line 3) is not important due to the low opacity but is likely to be modified due to the bright colouring.  The user therefore might want to prevent this word from being modified and instead give greater emphasis to words such as `boring' (line 4) that are important (high opacity) but are not likely to be chosen (dark colour).  This re-weighting of the evolutionary process can help the attack more quickly converge to better results.

\begin{figure*}[tb]
 \includegraphics[width=\textwidth]{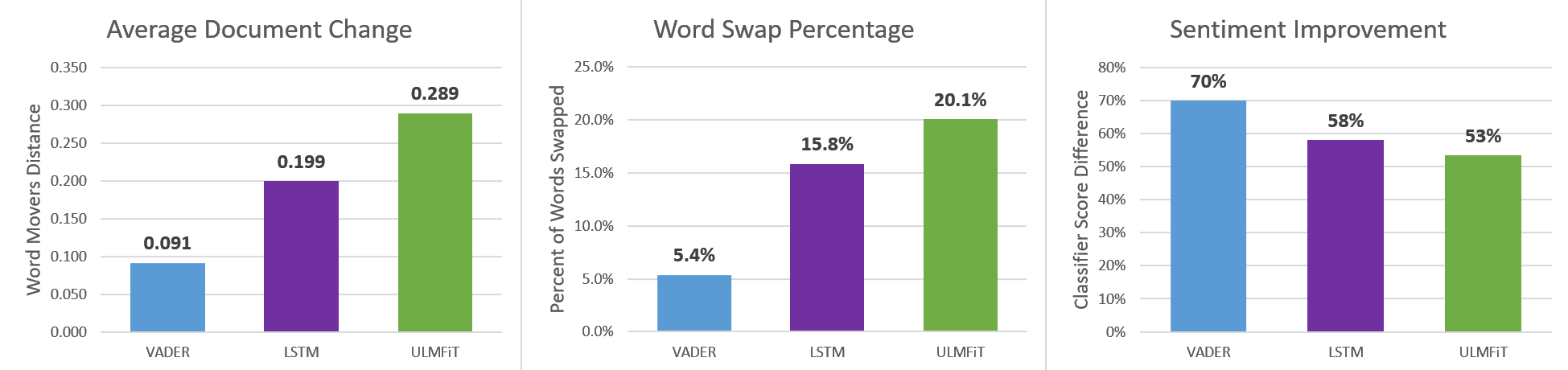}
 \caption{Robustness evaluation between classifiers on the IMDB dataset \cite{Maas} each attacked for 10 generations. Left: Average word mover's distance between original and adversarial documents. Center: Average percentage of word swapped between original and adversary.  Right: Average improvement in classifier score between original and adversary.}
  \vspace{-6mm}
 \label{fig:attack_results}
\end{figure*}

\subsection {Robustness Testing}
Another use of the framework is to compare the robustness of different classifiers. In this example the user tests three different classifiers by running the attack algorithm for each one and comparing the differences. The attack algorithm runs 10 generations for each record. The user is assessing how feasible it is for our attack to generate adversaries and to what extent each document is changed.

The data we use to demonstrate the attacks is the IMDB sentiment analysis dataset~\cite{Maas}.  This dataset has 50,000 movie reviews that are evenly split between negative and positive classes.  Each review is scored out of 10 by a human reviewer.  The review is negative if the score is less than or equal to 4 and positive if the score is greater than or equal to 7.  Neutral reviews are not available in the dataset. We test three different classifiers: VADER, a LSTM and ULMFiT.

VADER \cite{hutto2014vader} is a sentiment analysis classifier that uses a rules based approach built with a human curated lexicon of words. Each word placed on a spectrum from negative to positive. The LSTM is our implementation of an average performing deep learning classifier. The ULMFiT classifier \cite{Howard2018}  is a transfer learning method which was trained on the IMDB dataset \cite{Maas}.  As a baseline comparison between the models, we run the classifiers through the entire dataset without any adversarial testing.  VADER scores 59\%, the LSTM scores 84\% and ULMFiT scores 95.4\% (the highest accuracy of all existing works published on the dataset \cite{nlpprogress2019}).

As see in Figure~\ref{fig:attack_results}, for the results of the attack we measure word mover's distance, word swaps and sentiment improvement.  The word mover's distance is the difference between the final adversary at generation 10 and the original document.  The word swap is the percentage of words replaced between the final adversary and the original.  The sentiment improvement is the difference in classifier score between the original and final adversary. Scores from all classifiers are normalized in a range from -1.0 (100\% negative) to 1.0 (100\% positive) with a score of 0 considered a neutral score.  As an example a change from negative sentiment (-0.50) to positive (+0.25) would be a change of 75\%.  All the scores presented are the averages across all the records attacked. A more robust model will require a larger word mover's distance and more swapped words to reach the same classifier score as a weaker model.  A more robust model would also have a smaller classifier score improvement.

The ULMFiT classifier is the most robust because it had the largest word mover's distance and highest word swap percentage.  VADER had the least robust performance as very little perturbation needs to  be done in order to trick the classifier.  As little as 5\% of the words can be swapped with VADER compared to over 20\% for ULMFiT, the word mover's distance is also more than triple for ULMFiT (0.289) compared to VADER (0.091).  These results exemplify the importance of human edits for more robust models. Since a more robust model changes the documents more, there is a greater number of potential edits in need of fixing.  These tests also demonstrate that new attack strategies are needed to effectively attack the more complex models.  In addition to these robustness tests, another type of model assessment can be to compare different attack strategies against the same classifier in order to choose the best attack for further evaluation.

\section{Discussion and Future Work}
In this work we have presented a visual analytics framework that helps users build adversarial text examples.  While we have demonstrated that the framework can craft adversaries, there are many possible extensions and directions for future work.  Most importantly the system will undergo a more formal evaluation in which both quantitative and qualitative aspects of the work can be assessed. In this section we discuss some limitations and extensions of our approach as well as evaluation and other potential future work.

\subsection{Limitations}
The framework assumes we have unlimited access to the NLP classifier we are attacking.  This may not always be the case if for example, an online service has a maximum attempt lockout precaution or interprets our repeated queries as a denial of service attack. Mitigation techniques could include rate limiting our requests over time, a distributed attack, or slowly building a surrogate model that emulates the online system.  With a surrogate model made the attack can continue indefinitely in an offline setting.

The visual encodings used for the words are done by querying the classifier with each word to measure the influence of swapping each word.  When attacking a non-linear model, if any word is changed it can influence the results of any other subsequent changes.  Therefore each word must be reevaluated after any modification.  This becomes increasingly computationally expensive as we increase the number of words in the text.  Some methods to mitigate this issue could include filtering unimportant words, intelligent prefetching, or only encoding words upon user request. 

\subsection{Evaluation}
The robustness testing use case was a quantitative way of assessing our proposed framework.  However such calculations cannot be made as easily for more subjective matters such as text quality or model interpretability. This means that qualitative assessments in the form of user studies would also have to be done.  Methods of quantifying machine learning explanations have been considered in the evaluation of the XAI framework \cite{Aha2018}.  Methods have been suggested for developing a ``goodness scale'' using a 7-point Likert scale to assess factors such as plausibility, usefulness, clarity, causality, completeness, applicability, contrast and local as well as global observability.

For our future work we plan on conducting user studies through a method such as Amazon Mechanical Turk~\cite{mturk2019}.  Subjects would be provided samples of output from the attack algorithm after a set number of generations or the attack algorithm results plus edits made by a human reviewer.  The subjects would be asked to rate the semantic quality of the texts.  With ratings for both automation alone and human combined with machine we can compare the difference in ratings to assess the impact of human involvement in the generation process.

\subsection{Framework Extensions}
An extension to the views could include a filter to remove options such as different parts of speech, word tenses, or proper nouns.  This could let a user more quickly find a suitable word replacement. Another area for future work is the use of contextual word embeddings \cite{peters2018deep} that could provide more appropriate nearest neighbour by considering the local context of the word within the text. An extension to the current evolutionary algorithm can include a user-steerable stage of speculative execution~\cite{El-Assady2019}. This extension would track the quality of the text and will interrupt the process if a quality metric degrades past a certain threshold. At that point the system could present to the user various previews of new generations to allow the user to select the best path forward.

Other potential future work involves defensive measures for adversarial texts. The framework can be extended to test various defense strategies to help strengthen the models against adversarial examples. Most research on adversarial defense has been for computer vision and as discussed previously computer vision techniques do not often transition well to the discrete space of NLP.  Some recent works however have evaluated methods for adversaries using sequential data~\cite{rosenberg2019defense}.  These techniques were tested for cybersecurity and not NLP but their use of sequential methods could prove promising for NLP defense. As future work our system could test methods such as these and incorporate some auto machine learning techniques to search for optimal parameter settings. These suggestions can be added to the system for directing the user in choosing the best defensive measures against the attacks crafted by the user.

\section{Conclusion}
\added{In this work we propose a framework that extends existing evolutionary attack strategies to be used with a visual analytics dashboard.  As automated attack systems often degrade document semantics, the dashboard helps users make corrections to the texts. The framework is a black box and model-agnostic system so that it can work with any classifier that provides an output score. To start the user chooses an attack algorithm that perturbs each text over a set of generations which can be viewed in real time.  Once a set of adversaries has been built the user can make corrections manually or via suggestions by the system.  The suggestions are made by visually encoding the documents and providing replacement options in the scatterplot. The scatterplot displays choices based on a similarity score using word embeddings, semantic scores with a language model and the change to the classifer score. We demonstrate an implementation of the framework using a word swapping attack for sentiment analysis.  Use cases describe the process of making corrections to text semantics, troubleshooting an attack algorithm and evaluating the robustness of various classifiers.  This work is a first step towards further research integrating visual analytics with adversarial machine learning to encourage the exploration of robustness and interpretability techniques.}

\section*{Acknowledgements}
This research was supported by the Communications Security Establishment, the government of Canada's national cryptologic agency and the Natural Sciences and Engineering Research Council of Canada (NSERC).
\bibliographystyle{abbrv-doi}

\bibliography{main}
\end{document}